\documentclass[12pt]{article}
\usepackage{amsfonts}
\usepackage{eurosym}
\usepackage{graphicx}
\usepackage[utf8]{inputenc}
\usepackage[T1]{fontenc}
\usepackage{indentfirst}
\usepackage[margin=0.6in,nomarginpar]{geometry}
\usepackage[final]{hyperref}
\usepackage{amsmath}
\usepackage{hyperref}
\usepackage{cite}
\usepackage{subcaption}
\usepackage{caption}
\usepackage{amssymb}
\usepackage{multirow}
\usepackage[table]{xcolor}
\usepackage{orcidlink}

\hypersetup{
colorlinks=true,
linkcolor=blue,
citecolor=blue,
filecolor=magenta,
urlcolor=blue
}

\begin{document}

\title{One-dimensional Dunkl Quantum Mechanics: A Path Integral Approach}
\author{A. Benchikha\thanks{%
benchikha4@yahoo.fr} \\
$^{1}$D\'{e}partement de EC, Facult\'{e} de SNV, Universit\'{e} Constantine
1 Fr\`{e}res Mentouri,\\
Constantine, Algeria.\\
$^{2}$Laboratoire de Physique Math\'{e}matique et Subatomique,\\
LPMS, Universit\'{e}  Constantine 1 Fr\`{e}res Mentouri, Constantine,
Algeria. \and B. Hamil\thanks{%
hamilbilel@gmail.com(Corresponding author)} \\
Laboratoire de Physique Math\'{e}matique et Subatomique,\\
LPMS, Universit\'{e} Constantine 1 Fr\`{e}res Mentouri, Constantine,
Algeria. \and B. C. L\"{u}tf\"{u}o%
\u{g}lu\thanks{%
bekir.lutfuoglu@uhk.cz }, \\
Department of Physics, University of Hradec Kr\'{a}lov\'{e},\\
Rokitansk\'{e}ho 62, 500 03 Hradec Kr\'{a}lov\'{e}, Czechia.  \and B.Khantoul \thanks{%
boubakeur.khantoul@univ-constantine3.dz} \\
Department of Process Engineering, University of Constantine 3 - Salah
Boubnider, \\
BP: B72 Ali Mendjeli, 25000 Constantine, Algeria. }
\date{\today }
\maketitle

\begin{abstract}
In the present manuscript, we employ the Feynman path integral method to derive the propagator in one-dimensional Wigner-Dunkl quantum mechanics. To verify our findings we calculate the propagator associated with the free particle and the harmonic oscillator in the presence of the Dunkl derivative. We also deduce the energy spectra and the corresponding bound-state wave functions from the spectral decomposition of the propagator.
\end{abstract}

\section{Introduction}
Over the past decade, a new form of deformation in quantum mechanics has been explored by replacing the standard spatial derivative with the Dunkl derivative. The Dunkl derivative has its origins in a paper published in the mid-twentieth century in which Wigner addressed the question of whether quantum mechanical commutation relations could be derived from the classical equations of motion \cite{Wigner}. Yang subsequently revisited the same discussion for the simplest quantum mechanical system,  one-dimensional quantum harmonic oscillator, and demonstrated that a positive answer to Wigner's question is only possible by adding a reflection operator with a free parameter to the Heisenberg algebra, subject to various restrictive conditions \cite{Yang}. Subsequently, Green proposed a generalized field quantization method for bosons and fermions, utilizing the method of Wigner and Yang \cite{Green}. His findings introduced novel concepts to the fields of physics, designated as 'parastatistics, parabosons, parafermions', had played a pivotal role in the introduction of color degrees of freedom and quantum chromodynamics \cite{Greenberg}. Later, these concepts were subsequently examined extensively, particularly by Plyushchay in \cite{Mikn1, Mikn2, Mik1, Gamboa, Mik2, Klishevich}.

In 1989, in a different field of study, the pure mathematician Dunkl introduced a hybrid derivative by integrating the usual differential and parity operators with an arbitrary parameter for examining the relationships among the reflection groups and differential-difference operators \cite{Dunkl}. Interestingly, Dunkl's derivative, reminiscent of Yang's deformed momentum operator, eliminated some of the flaws \cite{heckman, deBie}. Subsequently, the Dunkl operator was used in deformed quantum mechanics together with the Calogero-Moser-Sutherland models \cite{Chakra, Lapointe, Kakei, Mikn3, Rodrigues, Horvathy, Mikn4}. Over the past decade, there has been a notable increase in interest surrounding the utilization of the Dunk derivative. The initial surge was initiated by the seminal papers of Genest et al., in which the isotropic, anisotropic, and supersymmetric Dunkl oscillators in the plane were examined \cite{Genest20131, Genest20132, Genest20133, Genest20141}. Subsequently, the three-dimensional solution for the isotropic Dunkl oscillator was provided in \cite{Genest20142}. Then, they demonstrated the superintegrability and exactly solvability of the Dunkl-Coulomb problem in the plane \cite{Vincent3}. Several years later, Ghazouani et al. generalized their solution to the three-dimensional case in \cite{Ghazouani2020}. In 2016, Isaac et al. classified the two-dimensional superintegrable Dunkl oscillator systems and proposed an algebraic derivation method to obtain their spectra \cite{Isaac2016}. Then, Salazar-Ram\'irez et al. considered Dunkl- oscillator and Coulomb problems to investigate the coherent states, respectively \cite{Salazar2017, Salazar2018}. In 2023, the authors constructed Dunkl-coherent states for the harmonic potential \cite{Ghazouani2023}, within the context of momentum representation \cite{ChungEPL2023}, and with the generalized form of the Dunkl derivative \cite{Sedaghatnia2023}. The impact of the Dunkl formalism on the relativistic equations has also been examined in many works. For example, in  \cite{Sargol2018} Sargolzaeipor et al. demonstrated its influence considering the Dirac equation in one dimension. Subsequently, Mota et al. obtained an exact analytical solution to the Dunkl-Dirac oscillator problem in two spatial dimensions in \cite{Mota20181}. Soon after,  they derived the energy spectrum of the one-dimensional Dunkl-Dirac oscillator with algebraic methods \cite{Ojeda2020}. In the last four years, different groups investigated the solutions for  Dunkl-  Klein-Gordon equation and oscillator\cite{Mota20211, Mota20212, Bilel20221}, Duffin-Kemmer-Petiau oscillator \cite{Merad2021, Merad2022} and  Pauli equation \cite{Bouguerne2024}. Additionally, the oscillators were treated as thermal systems and their thermodynamic properties were rigorously examined \cite{Dong2021, Bilel20222}. Some authors have proposed several generalizations of the Dunkl derivative \cite{Halberg2022, Samira2022, Mota20221, Dong2023, Rouabhia2023}. The Dunkl derivative is now used in many sub-fields, including noncommutative \cite{Samira2023} and non-constant curvatured \cite{Ballesteros2024} spaces, solid state physics \cite{Bilel20223}, quantum statistical mechanics \cite{Hassan2021, Fateh20231, Fateh20232, Hocine20241, Hocine20242} and others \cite{Junker2023, Quesne2023, Quesne2024, Schulze20241, Schulze20242, Schulze20243, Mota20241, Mota20242, Benzair2024, Junker}.  

Feynman's path integral formalism, introduced in his 1942 thesis \cite{Feynman}, presents an alternative to traditional quantum mechanics methods like Schrödinger and Heisenberg quantization. Feynman's method is notable for connecting classical and quantum physics by reintroducing a classical concept, trajectory, into quantum mechanics by establishing a connection between the classical Lagrangian description and quantum mechanics. The basic idea of path integration relies on the concept of the functional action, $S[x(t)]$, which in classical mechanics determines the unique path $x(t)$ that a particle takes between two points, $x_{a}$ and $x_{b}$. Unlike classical mechanics, there is no single path that describes the motion of a particle in quantum mechanics. Instead, the motion of a particle between two local points is characterized by a probability amplitude. Feynman demonstrated that this probability amplitude can be represented as a sum over all possible paths connecting to those certain points with a weight factor $exp(\frac{i}{h}S[x(t)])$. Recently, Junker \cite{Junker} constructed the path integral of Wigner-Dunkl quantum mechanics (WDQM) for even parity. He discovered that the path integral formalism in the WDQM is identical to the conventional Feynman path integral formalism of the standard quantum mechanics, with the only differences being the prefactor $\frac{1}{\vert x_{a}x_{b}\vert^{\nu }}$ and the effective potential $\frac{\nu }{x^{2}}.$ In this paper, with a simple change and using only algebraic calculations, we construct the path integral of the WDQM and show that the propagator is divided into two parity variants, even and odd.

This paper is organized as follows: in Section \ref{sec2}, we introduce the main relations of the one-dimensional WDQM. In Section \ref{sec3}, we follow the usual steps to derive the propagator for the one-dimensional Wigner quantum mechanics in the configuration space. Moreover, we give the spectral decomposition of the propagator. In Section \ref{sec4}, we calculate the propagator for the free particle and the harmonic oscillator in one dimension. Then, we extract the normalized wave functions and the corresponding energy eigenvalue functions. Finally, Section \ref{sec5} contains the conclusion.

\section{Quantum mechanics with Dunkl derivative}\label{sec2}

The WDQM, which we consider in this paper, is defined by the following commutation relation between the position and momentum operators,
\begin{equation}
\left[ \hat{x},\hat{p}\right] =i\hbar \left( 1+2\nu R\right) ,
\label{commutator}
\end{equation}%
where $\nu$ is the deformation parameter and is sometimes called the Wigner parameter in reference to historical connection. Here, $R$ denotes the reflection operator%
\begin{equation}
R=\left( -1\right) ^{x\partial _{x}}\text{, \ \ }Rf(x)=f(-x)\text{, \ \ }%
R^{2}=1.  \label{eigenvalu of reflection operator}
\end{equation}%
It is important to note that as $\nu \rightarrow 0$, the canonical
commutation relations of standard quantum mechanics are recovered. The representation of the deformed commutation relation, given in (\ref{commutator}), can be derived from operators $\hat{x}$ and $\hat{p}$, which satisfy the canonical commutation relations, by applying the transformation
\begin{equation}
\hat{p}=\dfrac{\hbar }{i}D_{x},\quad \hat{x}=x,  \label{Eq:1}
\end{equation}%
where $D_{x}$ is the Dunkl derivative, defined as 
\begin{equation}
D_{x}=\frac{d }{d x}+\dfrac{\nu }{x}(1-R).
\label{Dunkl derivative}
\end{equation}
This representation was thoroughly detailed for the first time by Dunkl in \cite{Dunkl}.  In this deformed algebra, the inner product in the Hilbert space associated with one-dimensional WDQM is defined by
\begin{equation}
\left\langle f\right\vert \left. g\right\rangle =\int_{-\infty }^{\infty
}f^{\ast }(x)g(x)|x|^{2\nu }dx,\text{ \ \ \ }\nu >-\frac{1}{2},
\end{equation}
where $|x|^{2\nu }$ is a weight function. Under these assumptions, the completeness and orthogonality relations can be extended as follows:
\begin{equation}
\int_{-\infty }^{\infty }dx|x|^{2\nu }\left\vert x\right\rangle \left\langle
x\right\vert =1,  \label{orthgo}
\end{equation}

\begin{equation}
\left\langle u\right\vert \left. x\right\rangle =\frac{1}{|x|^{2\nu }}\delta
\left( x-u\right) .  \label{delta}
\end{equation}%
Now, let us introduce a new momentum operator $\mathcal{\hat{P}}$ \cite{Grosche1, Grosche2, Grosche3, Grosche4} 
\begin{equation}
\mathcal{\hat{P}}=\frac{\hbar }{i}\left( \frac{\partial }{\partial x}+\frac{%
\nu }{x}\right) ,  \label{Newmomentum}
\end{equation}%
whose action on ket $\left\vert \mathcal{P}\right\rangle $ is%
\begin{equation}
\mathcal{\hat{P}}\left\vert \mathcal{P}\right\rangle =\mathcal{P}\left\vert 
\mathcal{P}\right\rangle \text{; \ \ }\left[ \hat{x},\mathcal{\hat{P}}\right]
=i\hbar .  \label{new commutation relation}
\end{equation}%
Then, on the coordinate space the eigenfunction for the new momentum operator, $\mathcal{\hat{P}}$, takes the following form 
\begin{equation}
\left\langle x\right\vert \mathcal{\hat{P}}\left\vert \mathcal{P}%
\right\rangle =\mathcal{P}\left\langle x\right\vert \left. \mathcal{P}%
\right\rangle =\frac{\hbar }{i}\left( \frac{\partial }{\partial x}+\frac{\nu 
}{x}\right) \left\langle x\right\vert \left. \mathcal{P}\right\rangle ,
\label{eq1}
\end{equation}%
and it can be solved to obtain formal momentum eigenvectors 
\begin{equation}
\left\langle \left\vert x\right\vert \right\vert \left. \mathcal{P}\right\rangle =\frac{1}{\sqrt{2\pi \hbar }}\left\vert x\right\vert ^{-\nu}e^{\frac{i}{\hbar }\mathcal{P}\text{ }\left\vert x\right\vert }=\frac{1}{\sqrt{2\pi \hbar }}y^{-\nu }e^{\frac{i}{\hbar }\mathcal{P}\text{ }y},\quad\text{with}\,\,\,\, y=\left\vert x\right\vert .  \label{newfermeturerelation}
\end{equation}%
In light of this, the identity operator for the eigenstates of the latter momentum operator reads:
\begin{equation}
\int_{-\infty }^{\infty }d\mathcal{P}\left\vert \mathcal{P}\right\rangle\left\langle \mathcal{P}\right\vert =1.  \label{momentumc}
\end{equation}%
We would like to emphasize that the objective of introducing the momentum operator, $\mathcal{\hat{P}}$, is to express the Dunkl-Hamiltonian in a familiar form that would allow us to investigate non-relativistic WDQM problems using the Feynman's path integral method.

\section{Propagator in one-dimensional Wigner-Dunkl quantum mechanics} \label{sec3}

In a non-relativistic quantum mechanical system, described by a Hamiltonian $\hat{H}$, the propagator or the transition amplitude from an initial state, $\left\vert x_{a}\right\rangle $, to a final state, $\left\vert x_{b}\right\rangle $, can be determined by the solution of the following equation
\begin{equation}
\left( \hat{H}_{b}-i\hbar \frac{\partial }{\partial t_{b}}\right) K(x_{b},t_{b};x_{a},t_{a})=i\hbar \delta \left( x_{b}-x_{a}\right) \delta \left( t_{b}-t_{a}\right) ,  \label{propa}
\end{equation}%
where the Dunkl-Hamilton operator is
\begin{equation}
\hat{H}=-\frac{\hbar ^{2}}{2m}\left[ \frac{\partial ^{2}}{\partial x^{2}}+%
\dfrac{2\nu }{x}\frac{\partial }{\partial x}-\frac{\nu }{x^{2}}\left(
1-R\right) \right] +V(\hat{x}).  \label{hamiltoniam}
\end{equation}%
The Dunkl-Hamiltonian operator of WDQM differs from the conventional quantum mechanical Hamiltonian by incorporating two additional terms. The first of these terms includes the first derivative with respect to the coordinate variable, while the second term depends on the reflection operator and distinguishes the odd parity solutions. 

Before formulating the propagator using Feynman path integrals in one-dimensional WDQM, we want to express the Dunkl-Hamiltonian in a form similar to that used in traditional quantum mechanics. For this purpose, using the Dunkl-momentum operator, we rewrite the Dunkl-Hamiltonian operator as follows : 
\begin{equation}
\hat{H}=\frac{\mathcal{\hat{P}}^{2}}{2m}+\frac{\hbar ^{2}\nu }{2mx^{2}}%
\left( \nu -R\right) +V(\hat{x}).  \label{Hamiltoniannew}
\end{equation}%
Subsequently, to eliminate the reflection operator, we introduce a regulation function
\begin{equation}
f\left( x\right) = f_{l}\left( x\right) f_{r}\left( x\right) ,    
\end{equation}
that specifies the parity 
\begin{equation}
Rf_{r}\left( x\right) =sf_{r}\left(x\right), \quad \text{with} \quad  s=\pm .
\end{equation}
Then, using the Duru-Kleinert method, we obtain a new form for the Hamiltonian
\begin{equation}
\hat{H}_{s}^{E}=f_{l}\left( x\right) \left( \hat{H}_{s}-E\right) f_{r}\left(x\right) ,
\end{equation}%
with the fixed-energy amplitude
\begin{equation}
G_{s}\left( x_{b},x_{a},E\right) =\left\langle x_{b}\left\vert \frac{1}{\hat{H}_{s}-E}\right\vert x_{a}\right\rangle =\frac{i}{\hbar }\int_{0}^{\infty}dTf_{r}\left( x_{b}\right) f_{l}\left( x_{a}\right) K_{s}^{E}(x_{b},x_{a};T),
\end{equation}%
where the Feynman kernel $K_{s}^{E}(x_{b},x_{a};T)$ is defined by \cite
{Kleinert}
\begin{equation}
K_{s}^{E}(x_{b},x_{a};T)=\left\langle x_{b}\right\vert e^{-\frac{i}{\hbar }%
\hat{H}_{s}^{E}T}\left\vert x_{a}\right\rangle .
\end{equation}
To obtain a path integral representation for $K_{s}^{E}(x_{b},x_{a};T)$, we use the standard method of discrediting the time interval $T$ into $N+1$ infinitesimal equal parts 
\begin{equation}
\varepsilon =\frac{T}{N+1}=\varepsilon_{s}f_{l}\left( x_{j}\right) f_{r}\left( x_{-1}\right).     
\end{equation}
We then apply the Trotter's formula, $e^{-\frac{i}{\hbar}\hat{H}_{s}^{E}T}=\left[ e^{-\frac{i}{\hbar }\varepsilon \hat{H}_{\nu }^{E}}\right] ^{N+1}$ and insert the closure relations (\ref{orthgo}) and (\ref{momentumc}) between each pair of exponential in order to eliminate the operators $\hat{x}$ and $\mathcal{\hat{P}}$. With the help of the scalar product 
\begin{equation}
\left\langle \left\vert x\right\vert \right\vert \left. \mathcal{P}\right\rangle =\frac{1}{\sqrt{2\pi \hbar }}\left\vert x\right\vert ^{-\nu }e^{\frac{i}{\hbar }\mathcal{P}\text{ }\left\vert x\right\vert },
\end{equation}
which allows it to pass from one base to another one,  we obtain the kernel, $K_{E}^{\nu }$, in the following discrete form,
\begin{eqnarray}
K_{s}^{E}(y_{b},y_{a};T) &=&\underset{N\rightarrow \infty }{\lim }\int \overset{N}{\underset{j=1}{\prod }}y_{j}^{2\nu }dy_{j}\overset{N+1}{\underset{n=1}{\prod }}\frac{d\mathcal{P}_{j}}{2\pi \hbar \left(y_{j}y_{j-1}\right) ^{\nu }}\exp \text{ }\left\{ \frac{i}{\hbar }\overset{N}{\underset{j=1}{\sum }}\Bigg[ \mathcal{P}_{j}\left( y_{j}-y_{j-1}\right)\right.  \notag \\
&&. \left. -\varepsilon _{s}f_{r}\left( y_{j}\right) f_{l}\left(y_{j-1}\right) \left( \frac{\mathcal{P}_{j}^{2}}{2m}+\frac{\hbar ^{2}\left(\lambda _{s}^{2}-\frac{1}{4}\right) }{y_{j}^{2}}+V\left( y_{j}\right)-E\right) \Bigg] \right\} ,  \label{mesurekernel}
\end{eqnarray}%
where $y_{a}=y_{0}$, $y_{b}=y_{N+1}$, $t_{a}=t_{0}$, $t_{b}=t_{N+1}$ and
\begin{equation}
\lambda _{s}=\nu -\frac{s}{2}.  \label{gamma}
\end{equation}%
By establishing a symmetrical measure, 
\begin{equation}
  \overset{N}{\underset{j=1}{\prod }}y_{j}^{2\nu
}=\frac{1}{\left( y_{a}y_{b}\right) ^{\nu }}\overset{N+1}{\underset{j=1}{%
\prod }}\left( y_{j}y_{j-1}\right) ^{\nu },
\end{equation}
the partial kernel then becomes 
\begin{eqnarray}
K_{s}^{E} &=&\underset{N\rightarrow \infty }{\lim }\frac{1}{\left(
y_{a}y_{b}\right) ^{\nu }}\int \overset{N}{\underset{j=1}{\prod }}dy_{j}%
\overset{N+1}{\underset{n=1}{\prod }}\frac{d\mathcal{P}_{j}}{2\pi \hbar }%
\exp \text{ }\left\{ \frac{i}{\hbar }\overset{N}{\underset{j=1}{\sum }}\Bigg[
\mathcal{P}_{j}\left( y_{j}-y_{j-1}\right) \right.   \notag \\
&& \left. -\varepsilon _{s}f_{l}\left( y_{j}\right) f_{r}\left(
y_{j-1}\right) \left( \frac{\mathcal{P}_{j}^{2}}{2m}+\frac{\hbar ^{2}\left(
\lambda _{s}^{2}-\frac{1}{4}\right) }{y_{j}^{2}}+V\left( y_{j}\right)
-E\right) \Bigg] \right\} ,
\end{eqnarray}%
After evaluating the integrals with respect to $\mathcal{P}_{j}$, we arrive at 
\begin{eqnarray}
K_{s}^{E} &=&\underset{N\rightarrow \infty }{\lim }\frac{1}{\left(
y_{a}y_{b}\right) ^{\nu }}\sqrt{\frac{m}{2\pi i\varepsilon _{s}f_{l}\left(
y_{b}\right) f_{r}\left( y_{a}\right) \hbar }}\int \overset{N}{\underset{j=1}%
{\prod }}\left[ \sqrt{\frac{m}{2\pi i\varepsilon _{s}f\left( y_{j}\right)
\hbar }}dy_{j}\right]  \notag \\
&&\times \exp \left\{ \frac{i}{\hbar }\overset{N}{\underset{j=1}{\sum }}%
\left[ \frac{m\left( \Delta y_{j}\right) ^{2}}{2\varepsilon _{s}f_{l}\left(
y_{j}\right) f_{r}\left( y_{j-1}\right) }\right.  \left. -\varepsilon _{s}f_{l}\left( y_{j}\right) f_{r}\left(
y_{j-1}\right) \left( \frac{\hbar ^{2}\left( \lambda _{s}^{2}-\frac{1}{4}%
\right) }{y_{j}^{2}}+V\left( y_{j}\right) -E\right) \right] \right\} .
\label{kdiscret}
\end{eqnarray}%
As long as the condition $f\left( y\right) =f_{l}\left( y\right) f_{r}\left(
y\right) $ is satisfied, the functions $f_{l}\left( y\right) $ and $%
f_{r}\left( y\right) $ can be chosen arbitrarily \cite{Kleinert}. Thus, we
can adopt Kleinert's approach by assuming $f_{l}\left( y\right) \equiv
f_{r}\left( y\right) \equiv 1.$ Then, the Feynman kernel $K_{E}^{\nu}(x_{b},x_{a};T)$ reads: 
\begin{equation}
K_{s}^{E}(y_{b},y_{a};T)=e^{\frac{i}{\hbar }ET}K_{s}(y_{b},y_{a};T), \label{propagatorfinal 1}
\end{equation}%
with
\begin{equation}
K_{s}(y_{b},y_{a};\lambda )=\frac{1}{\left( y_{a}y_{b}\right) ^{\nu }}\int Dy\exp \left\{ \frac{i}{\hbar }\int_{0}^{T}dt\left[ \frac{m}{2}\dot{y}^{2}-V_{eff}(y)\right] \right\} ,  \label{propagatorfinal}
\end{equation}%
and the effective potential
\begin{equation}
V_{eff}\left( y\right) =\frac{\hbar ^{2}\left( \lambda _{s}^{2}-\frac{1}{4}\right) }{2my^{2}}+V\left( y\right) .  \label{effectivepotential}
\end{equation}%
Here, we note that the obtained kernel, given in equation (\ref{propagatorfinal}), is the same as the conventional Feynman path integral of the standard quantum mechanics, except with two terms. These terms are the prefactor ${\left( y_{a}y_{b}\right) ^{-\nu }}$ and the additional repulsive singular potential term, $\frac{\hbar ^{2} }{2my^{2}}\left( \lambda _{s}^{2}-\frac{1}{4}\right)$, which exists in the effective potential. We have to underline the fact that the effective potential, given in   (\ref{propagatorfinal}), differs from Junker's findings  \cite{Junker}. To be more precise,  in the current form, the kernel propagator is split into two parity variants,  even and odd, designated by the eigenvalues $s$. Here, the even parity case propagator%
\begin{equation}
K_{+}(y_{b},y_{a};T)=\frac{1}{\left( y_{a}y_{b}\right) ^{\nu }}\int Dy\exp
\left\{ \frac{i}{\hbar }\int_{0}^{T}dt\left[ \frac{m}{2}\dot{y}^{2}-\frac{%
\hbar ^{2}\left( \alpha ^{2}-\frac{1}{4}\right) }{2my^{2}}-V(y)\right]
\right\} ,  \label{k+}
\end{equation}
and the odd parity case propagator 
\begin{equation}
K_{-}(y_{b},y_{a};T)=\frac{1}{\left( y_{a}y_{b}\right) ^{\nu }}\int Dy\exp
\left\{ \frac{i}{\hbar }\int_{0}^{T}dt\left[ \frac{m}{2}\dot{y}^{2}-\frac{%
\hbar ^{2}\left( \beta ^{2}-\frac{1}{4}\right) }{2my^{2}}-V(y)\right]
\right\} ,  \label{k-}
\end{equation}%
stand with
\begin{equation}
\alpha =\nu -\frac{1}{2},\text{ \ \ }\beta =\nu +\frac{1}{2}.
\label{deformedparameters}
\end{equation}
Taking into account (\ref{k+}) and (\ref{k-}), we can write the Dunkl path integral kernel as a sum of even and odd kernel propagators as follows \cite{Junker}:
\begin{equation}
K(y_{b},y_{a};T)=K_{+}(y_{b},y_{a};T)+sgn\left( y_{b}y_{a}\right)K_{-}(y_{b},y_{a};T),  \label{propagatoroe}
\end{equation}%
where $sgn\left( x\right) $ represents the sign function. Please note that if the Hamiltonian (\ref{hamiltoniam}) has a complete set of eigenfunctions $\Psi_{n,s}$ associated with eigenvalues $E_{n}^{s}$, we then can apply the following spectral decomposition%
\begin{equation}
K(y_{b},y_{a};T)=\underset{n}{\sum }\left\{ \left[ \Psi _{n,+1}\left(y_{b}\right) \Psi _{n,+1}^{^{\ast }}\left( y_{a}\right) \right] e^{-\frac{i}{\hbar }E_{n}^{+}T}+\left[ \Psi _{n,-1}\left( y_{b}\right) \Psi_{n,-1}^{^{\ast }}\left( y_{a}\right) \right] e^{-\frac{i}{\hbar }E_{n}^{-}T}\right\} ,  \label{spectral decomposition 2}
\end{equation}%
where $E_{n,\nu }^{+}$ and $E_{n,\nu }^{-}$ are the energy eigenvalues related to the even and odd wave functions, respectively.

\section{Applications} \label{sec4}
To demonstrate the efficacy of our method, we apply it to two fundamental problems: the free particle case and the harmonic oscillator. 

\subsection{Free particle} \label{subsec1}
This subsection aims to derive the Dunkl kernel propagator for the free particle scenario.  In one dimension, with $V\left( x\right) =0$, we express the Dunkl kernel propagator via \eqref{propagatoroe} as follows:
\begin{eqnarray}
K(x_{b},x_{a};T) &=&\frac{1}{\left\vert x_{a}x_{b}\right\vert ^{\nu }}\int D\left\vert x\right\vert \exp \left\{ \frac{i}{\hbar }\int_{0}^{T}dt\left[ \frac{m}{2}\dot{x}^{2}-\frac{\hbar ^{2}\left(z\alpha ^{2}-\frac{1}{4}\right) }{2mx^{2}}\right] \right\}  \notag \\
&&+\frac{1}{\left\vert x_{a}x_{b}\right\vert ^{\nu }}sng\left( x_{a}x_{b}\right) \int D\left\vert x\right\vert \exp \left\{ \frac{i}{\hbar }\int_{0}^{T}dt\left[ \frac{m}{2}\dot{x}^{2}-\frac{\hbar ^{2}\left( \beta^{2}-\frac{1}{4}\right) }{2mx^{2}}\right] \right\} .  \label{341}
\end{eqnarray}%
We observe that the latter propagator is identical to the propagator of a free particle moving in two dimensions, as referenced in \cite{Grosche2}. Consequently, we express the final form of the  propagator as follows: 
\begin{eqnarray}
K(x_{b},x_{a};T) &=&\frac{1}{\left\vert x_{a}x_{b}\right\vert ^{\nu -\frac{1%
}{2}}}\frac{m}{2i\hbar T}\exp \left[ \frac{im}{2\hbar T}\left(
x_{a}^{2}+x_{b}^{2}\right) \right]  \notag \\
&&\times \left\{ I_{\nu -\frac{1}{2}}\left( \frac{m\left\vert
x_{a}x_{b}\right\vert }{i\hbar T}\right) +sng\left( x_{a}x_{b}\right) I_{\nu
+\frac{1}{2}}\left( \frac{m\left\vert x_{a}x_{b}\right\vert }{i\hbar T}%
\right) \right\} , 
\end{eqnarray}
where $I_{\mu }\left( x\right) $ is the modified Bessel functions of the first kind. Then, we use the relation between $I_{\mu }\left( x\right) $ and deformed exponential function $E_{\mu }\left( x\right) $ ,
\begin{equation}
E_{\nu }\left( x\right) =\Gamma \left( \nu +\frac{1}{2}\right) \left( \frac{2}{\left\vert x\right\vert }\right) ^{\nu -\frac{1}{2}}\left( I_{\nu -\frac{1}{2}}\left( \left\vert x\right\vert \right) +sng\left( x\right) I_{\nu +\frac{1}{2}}\left( \left\vert x\right\vert \right) \right) ,
\label{deformedexponential}
\end{equation}%
which results in
\begin{equation}
K(x_{b},x_{a};T)=\frac{1}{\Gamma \left( 2\nu +1\right) }\left( \frac{m}{2i\hbar T}\right) ^{\nu +\frac{1}{2}}e^{\frac{im}{2\hbar T}\left(x_{a}^{2}+x_{b}^{2}\right) }E_{\nu }\left( \frac{mx_{a}x_{b}}{i\hbar T}%
\right) . \label{can1}
\end{equation}
It is important to underline that \eqref{can1} is the same as equation (52) in \cite{Junker}.

\subsection{Harmonic oscillator}\label{subsec2}

We now examine the harmonic potential problem. In this case, with $V\left(x\right) =\frac{1}{2}m\omega ^{2}x^{2}$, the kernel propagator reduces to 
\begin{eqnarray}
K(x_{b},x_{a};T) &=&\frac{1}{\left\vert x_{a}x_{b}\right\vert ^{\nu }}\int D\left\vert x\right\vert \exp \left\{ \frac{i}{\hbar }\int_{0}^{T}dt\left[ \frac{m}{2}\dot{x}^{2}-\frac{1}{2}m\omega ^{2}x^{2}-\frac{\hbar ^{2}\left( \alpha ^{2}-\frac{1}{4}\right) }{2mx^{2}}\right] \right\}  \notag \\
&&+\frac{1}{\left\vert x_{a}x_{b}\right\vert ^{\nu }}sng\left( x_{a}x_{b}\right) \int D\left\vert x\right\vert \exp \left\{ \frac{i}{\hbar } \int_{0}^{T}dt\left[ \frac{m}{2}\dot{x}^{2}-\frac{1}{2}m\omega ^{2}x^{2}-\frac{\hbar ^{2}\left( \beta ^{2}-\frac{1}{4}\right) }{2mx^{2}}\right]\right\} .  \,\,\,\,
\end{eqnarray}
This corresponds in fact to the d-dimensional isotropic harmonic oscillator that is perturbed by an inverse square potential. Using the Hille-Hardy formula 
\begin{equation}
\underset{n=0}{\overset{\infty }{\sum }}\frac{n!L_{n}^{\mu }\left( x\right) L_{n}^{\mu }\left( y\right) z^{n}}{\Gamma \left( n+\mu +1\right) }e^{-\frac{x+y}{2}}=\frac{\left( xyz\right) ^{-\mu /2}}{\left( 1-z\right) }\exp \left[ - \frac{1}{2}\left( x+y\right) \frac{1+z}{1-z}\right] I_{\mu }\left( \frac{2\sqrt{xyz}}{1-z}\right) ,
\end{equation}%
we obtain a spectral decomposition of the kernel propagators
\begin{eqnarray}
K(x_{b},x_{a};T) &=&\frac{1}{\left\vert x_{a}x_{b}\right\vert ^{^{\nu -\frac{1}{2}}}}\frac{m\omega }{\hbar }\underset{n}{\sum }\frac{n!}{\Gamma \left(n+\nu +\frac{1}{2}\right) }\left( \frac{m\omega \left\vert x_{a}x_{b}\right\vert }{\hbar }\right) ^{\nu -\frac{1}{2}}  \notag \\
&&\times L_{n}^{\nu -\frac{1}{2}}\left( \frac{m\omega }{\hbar }x_{a}^{2}\right) L_{n}^{\nu -\frac{1}{2}}\left( \frac{m\omega }{\hbar } x_{b}^{2}\right) e^{-\frac{m\omega }{2\hbar }\left( x_{a}^{2}+x_{b}^{2}\right) -i\omega T\left( 2n+\nu +\frac{1}{2}\right) } \notag \\
&&+sng\left( x_{a}x_{b}\right) \frac{1}{\left\vert x_{a}x_{b}\right\vert ^{^{\nu -\frac{1}{2}}}}\frac{m\omega }{\hbar }\underset{n}{\sum }\frac{n!}{\Gamma \left( n+\nu +\frac{3}{2}\right) }\left( \frac{m\omega \left\vert x_{a}x_{b}\right\vert }{\hbar }\right) ^{\nu +\frac{1}{2}}  \notag \\
&&\times L_{n}^{\nu +\frac{1}{2}}\left( \frac{m\omega }{\hbar } x_{a}^{2}\right) L_{n}^{\nu +\frac{1}{2}}\left( \frac{m\omega }{\hbar } x_{b}^{2}\right) e^{-\frac{m\omega }{2\hbar }\left(x_{a}^{2}+x_{b}^{2}\right) -i\omega T\left( 2n+\nu +\frac{3}{2}\right) }.
\end{eqnarray}%
After the algebra, we obtain the normalized wave functions and their energy eigenvalues, that describe the one-dimensional quantum harmonic oscillator parity-dependent dynamics as follows: 
\begin{itemize}
\item \textbf{\ }Even parity solution
\end{itemize}

\begin{equation}
\Psi _{n,\nu }^{+}\left( x\right) =\sqrt{\frac{n!}{\Gamma \left( n+\nu +%
\frac{1}{2}\right) }}\left( \frac{m\omega }{\hbar }\right) ^{\frac{\nu }{2}+%
\frac{1}{4}}\exp \left[ -\frac{m\omega }{2\hbar }x^{2}\right] L_{n}^{\nu -%
\frac{1}{2}}\left( \frac{m\omega }{\hbar }x^{2}\right) .
\end{equation}

\begin{equation}
E_{n}^{+}=\hbar \omega \left( 2n+\nu +\frac{1}{2}\right) ,\text{ \ \ \ }\ \
n=0,1,2,...
\end{equation}

\begin{itemize}
\item Odd parity solution
\end{itemize}

\bigskip

\begin{equation}
\Psi _{n}^{-}\left( x\right) =\sqrt{\frac{n!}{\Gamma \left( n+\nu +\frac{3}{2%
}\right) }}\text{ }\left( \frac{m\omega }{\hbar }\right) ^{\frac{\nu }{2}+%
\frac{3}{4}}x\exp \left[ -\frac{m\omega }{2\hbar }x^{2}\right] L_{n}^{\nu +%
\frac{1}{2}}\left( \frac{m\omega }{\hbar }x^{2}\right) .
\end{equation}%
\begin{equation}
E_{n}^{-}=\hbar \omega \left( 2n+\nu +\frac{3}{2}\right) ,\text{ \ \ \ \ \ \ 
}n=0,1,2,...
\end{equation}

\section{Conclusion}\label{sec5}


In this work, we have constructed the propagator of one-dimensional quantum systems with Dunkl derivative using the configuration space representation and the Feynman path integral approach. We have found that the resulting form of the propagator contains two new terms. The first one of these terms arises in the measure and depends on the free parameter as ${\left\vert x_{a}x_{b}\right\vert ^{-\nu }}$. The second term is a radial term of the form $\frac{\hbar ^{2} }{2my^{2}}\left( \lambda _{s}^{2}-\frac{1}{4}\right)$, and stands in the effective potential. We note that both new terms disappear as the free parameter approaches zero. Then, we validated our results in two well-known examples, the free particle and the quantum harmonic potential cases. In the free particle case, we obtained the wave function and compared it with the literature. In the harmonic oscillator problem, we found the energy eigenvalue function in addition to the wave function for even and odd parity cases. We conclude that our methodology can be used efficiently to study other problems of physics, and may be extended within higher dimensions using the polar and the spherical coordinates.

\section*{Acknowledgments}

This work is supported by the Ministry of Higher Education and Scientific Research, Algeria under the code: B00L02UN040120230003.


\begin{thebibliography}{99}
\bibitem{Wigner} E. P. Wigner, Phys. Rev. \textbf{77}, 711 (1950).

\bibitem{Yang} L. M. Yang, Phys. Rev. \textbf{84}, 788 (1951).

\bibitem{Green} H. S. Green, Phys. Rev. \textbf{90}, 270 (1953).

\bibitem{Greenberg} O. W. Greenberg, Phys. Rev. Lett. \textbf{13}, 598 (1964).

\bibitem{Mikn1} M. S. Plyushchay, Phys. Lett. B \textbf{320}, 91 (1994). 

\bibitem{Mikn2} M. S. Plyushchay, Ann. Phys. \textbf{245}, 339 (1996). 

\bibitem{Mik1} M. Plyushchay, Nucl. Phys. B \textbf{491}, 619 (1997).

\bibitem{Gamboa} J. Gamboa, M. Plyushchay, J. Zanelli, Nucl. Phys. B \textbf{543}, 447 (1999). 

\bibitem{Mik2} M. Plyushchay, Int. J. Mod. Phys. A \textbf{15}, 3679 (2000).

\bibitem{Klishevich} S. M. Klishevich, M. Plyushchay, M. Rausch de Traubenberg, Nucl. Phys. B \textbf{616}, 419 (2001).


\bibitem{Dunkl} C. F. Dunkl, Trans. Am. Math. Soc. \textbf{311}(1), 167 (1989).


\bibitem{heckman} G. J. Heckman, Prog. Math. \textbf{101}, 181 (1991).

\bibitem{deBie} H. de Bie, B. Orsted, P. Somberg, V. Sou\v{c}ek, Trans.
Am. Math. Soc. \textbf{364}, 3875 (2012).



\bibitem{Chakra} R. Chakrabarti, R. Jagannathan, J. Phys. A: Math Gen \textbf{27}, L227 (1994).

\bibitem{Lapointe} L. Lapointe, L. Vinet, Commun. Math. Phys. \textbf{178}(2), 425 (1996).

\bibitem{Kakei} S. Kakei, J. Phys. A \textbf{29}, L619 (1996).

\bibitem{Mikn3} P. A. Hortv\'athy, M. S. Plyushchay, Phys. Lett. B \textbf{595}, 547 (2004). 

\bibitem{Rodrigues} R. de Lima Rodrigues, J. Phys.  A \textbf{42}, 355213 (2009).

\bibitem{Horvathy} P. A. Hortv\'athy, M. Plyushchay, M. Valenzuela, Ann. Phys. \textbf{325}, 1931 (2010).

\bibitem{Mikn4} F. Correa, O. Lechtenfeld, M. S. Plyushchay, J. High Energy Phys. \textbf{2014}, 151 (2014). 


\bibitem{Genest20131} V. X. Genest, M. E. H. Ismail, L. Vinet, A. Zhedanov, J. Phys. A: Math. Theor. \textbf{46}, 145201 (2013).

\bibitem{Genest20132} V. X. Genest, L. Vinet, A. Zhedanov, J. Phys. A: Math. Theor. \textbf{46}, 325201 (2013). 

\bibitem{Genest20133} V. X. Genest, J. M. Lemay, L. Vinet, A. Zhedanov, J. Phys. A: Math. Theor. \textbf{46}, 505204 (2013). 

\bibitem{Genest20141} V. X. Genest, M. E. H. Ismail, L. Vinet, A. Zhedanov, Commun. Math. Phys. \textbf{329}, 999 (2014).

\bibitem{Genest20142} V. X. Genest L. Vinet, A. Zhedanov, J. Phys.: Conf. Ser. \textbf{512}, 012010 (2014).

\bibitem{Vincent3} V. X. Genest, Andr\'{e}anne Lapointe, Luc Vinet,
Phys. Lett. A, \textbf{379}, 923 (2015).

\bibitem{Ghazouani2020} S. Ghazouani, S. Insaf, J. Phys. A: Math. Theor. \textbf{46}, 035202 (2020). 

\bibitem{Isaac2016} P. S. Isaac, I. Marquette, J. Phys. A: Math. Theor. \textbf{49}, 115201 (2016). 

\bibitem{Salazar2017} M. Salazar-Ram\'irez, D. Ojeda-Guill\'en, R. D. Mota, V. D. Granados, Eur. Phys. J. Plus \textbf{132}, 39 (2017).

\bibitem{Salazar2018} M. Salazar-Ram\'irez, D. Ojeda-Guill\'en, R. D. Mota, V. D. Granados, Mod. Phys. Lett. A \textbf{33}, 1850112 (2018).


\bibitem{Ghazouani2023} S. Ghazouani, J. Phys. A: Math. Theor. \textbf{55}, 505203 (2023).

\bibitem{ChungEPL2023} W. S. Chung, M. de Montigny, H. Hassanabadi, EPL \textbf{141}, 60004 (2023). 

\bibitem{Sedaghatnia2023} P. Sedaghatnia, H. Hassanabadi, G. Junker, J. K\u{r}\'i\u{z}, W. S. Chung, Ann. Phys. \textbf{458}, 169445 (2023).

\bibitem{Sargol2018} S. Sargolzaeipor, H. Hassanabadi, W. S. Chung, Mod. Phys. Lett. A \textbf{33}, 1850146 (2018).


\bibitem{Mota20181} R. D. Mota, D. Ojeda-Guill\'{e}n, M. Salazar-Ram\'{\i}rez, V. D. Granados, Ann. Phys. \textbf{411}, 167964 (2019).

\bibitem{Ojeda2020} D. Ojeda-Guill\'{e}n, R. D. Mota, M. Salazar-Ram\'{\i}rez, V. D. Granados, Mod. Phys. Lett. A \textbf{35}, 2050255 (2020).



\bibitem{Mota20211} R. D. Mota, D. Ojeda-Guill\'{e}n, M. Salazar-Ramirez, V. D. Granados, Mod. Phys. Lett. A \textbf{36},  2150066
(2021).

\bibitem{Mota20212} R. D. Mota, D. Ojeda-Guill\'{e}n, M. Salazar-Ramirez, V. D. Granados, Mod. Phys. Lett. A \textbf{36},  2150171
(2021).

\bibitem{Bilel20221} B. Hamil, B. C. L\"{u}tf\"{u}o\u{g}lu, Few-Body Syst. \textbf{63}, 74 (2022).

\bibitem{Merad2021} A. Merad, M. Merad, Few-Body Syst. \textbf{62}, 98 (2021).

\bibitem{Merad2022} A. Merad, M. Merad, T. Boudjedaa, Int. J. Mod. Phys. A \textbf{37}, 2250072 (2022).

\bibitem{Bouguerne2024} H. Bouguerne, B. Hamil, B. C. L\"{u}tf\"{u}o\u{g}lu, M. Merad,  \href{https://doi.org/10.1007/s12648-024-03170-y}{Ind. J. Phys. \textbf{Early access},  (2024).}


\bibitem{Dong2021} S. H. Dong, W. H. Huang, W. S. Chung, P. Sedaghatnia, H. Hassanabadi, EPL \textbf{135}, 30006 (2021).

\bibitem{Bilel20222} B. Hamil, B. C. L\"{u}tf\"{u}o\u{g}lu, Eur. Phys. J. Plus \textbf{137}, 812 (2022).

\bibitem{Halberg2022} A. Schulze-Halberg, Phys. Scr. \textbf{97}, 085213 (2022).

\bibitem{Samira2022} S. Hassanabadi, J.K\v{r}\'{\i}\v{z}, B. C. L\"{u}tf\"{u}o\u{g}lu, H. Hassanabadi, Phys. Scr. \textbf{97}, 125305 (2022).


\bibitem{Mota20221} R. D. Mota, D. Ojeda-Guill\'{e}n,  Mod. Phys. Lett. A \textbf{37},  2250224
(2022).

\bibitem{Dong2023} S. H. Dong, L. F. Quezada, W. S. Chung, P. Sedaghatnia, H. Hassanabadi, Ann. Phys. \textbf{451}, 169259 (2023).

\bibitem{Rouabhia2023} N. Rouabhia, M. Merad, B. Hamil, EPL \textbf{143}, 52003 (2023).

\bibitem{Samira2023} S. Hassanabadi, P. Sedaghatnia, W. S. Chung, B. C. L\"{u}tf\"{u}o\u{g}lu, J.K\v{r}\'{\i}\v{z},  H. Hassanabadi, Eur. Phys. J. Plus \textbf{138}, 331 (2023).

\bibitem{Ballesteros2024} A. Ballesteros, A. Najafizade, H. Panahi, H. Hassanabadi, S. H. Dong, Ann. Phys. \textbf{460}, 169543 (2024).



\bibitem{Bilel20223} B. Hamil, B. C. L\"{u}tf\"{u}o\u{g}lu, Eur. Phys. J. Plus \textbf{137}, 1241 (2022).

\bibitem{Hassan2021} H. Hassanabadi, M. de Montigny, W. S. Chung, Physica A \textbf{580}, 126154 (2021).

\bibitem{Fateh20231} F. Merabtine, B. Hamil, B. C. L\"{u}tf\"{u}o\u{g}lu, A. Hocine, M. Benarous, J. Stat. Mech.  \textbf{2023}, 053102 (2023). 

\bibitem{Fateh20232} F. Merabtine, B. Hamil, B. C. L\"{u}tf\"{u}o\u{g}lu, Physica A \textbf{623}, 128841 (2023). 

\bibitem{Hocine20241} A. Hocine, B. Hamil, F. Merabtine,  B. C. L\"{u}tf\"{u}o\u{g}lu,  M. Benarous, Rev. Phys. Mex.  \textbf{in press},  (2024).

\bibitem{Hocine20242} A. Hocine, F. Merabtine, B. Hamil, B. C. L\"{u}tf\"{u}o\u{g}lu, M. Benarous,  \href{https://doi.org/10.1007/s12648-024-03311-3}{Ind. J. Phys. \textbf{Early access},  (2024).}


\bibitem{Junker2023} G. Junker, S. H. Dong, P. Sedaghatnia, W. S. Chung, H. Hassanabadi, Ann. Phys. \textbf{454}, 169336 (2023).

\bibitem{Quesne2023} C. Quesne J. Phys. A: Math. Theor. \textbf{56}, 265203 (2023).

\bibitem{Quesne2024} C. Quesne, EPL \textbf{145}, 62001 (2024).

\bibitem{Mota20241} R. D. Mota, D. Ojeda-Guill\'{e}n,  M. A. Xicot\'encatl, Physica A \textbf{635},  129525
(2024).

\bibitem{Mota20242} R. D. Mota, D. Ojeda-Guill\'{e}n,  M. A. Xicot\'encatl, Few-Body Syst. \textbf{65},  25
(2024).

\bibitem{Schulze20241} A. Schulze-Halberg, P. Roy, J. Phys. A \textbf{57}, 225204 (2024).

\bibitem{Schulze20242} A. Schulze-Halberg, Few-Body Syst. \textbf{65}, 58 (2024).

\bibitem{Schulze20243} A. Schulze-Halberg, Int. J. Mod. Phys. A \textbf{39}, 2450013 (2024).


\bibitem{Benzair2024} H. Benzair, T. Boudjedaa, M. Merad, Phys. Scr. \textbf{99}, 055261 (2024). 

\bibitem{Junker} G. Junker, J. Phys. A: Math. Theor. \textbf{57}, 075201 (2024).


\bibitem{Feynman} R. P. Feynman, A. Hibbs, \emph{Quantum Mechanics and Path Integrals}, (McGraw Hill, New York), 1965.


\bibitem{Grosche1} C. Grosche, G. S. Pogosyan, A. N. Sissakian, Fortschr. Phys. \textbf{43}, 453 (1995).

\bibitem{Grosche2} C. Grosche, F. Steiner,  \emph{Handbook of Feynman Path Integrals. Springer Tracts in Modern Physics 145}, (Springer, Berlin - Heidelberg), 1998.

\bibitem{Grosche3} C. Grosche, Phys. Part. Nuclei \textbf{37}, 368 (2006).

\bibitem{Grosche4} C. Grosche, G. S. Pogosyan, A. N. Sissakian, Phys. Part. Nuclei \textbf{38}, 299 (2007).

\bibitem{Kleinert} H. Kleinert, \emph{Path Integrals in Quantum Mechanics, Statistics, Polymer Physics and Financial Markets}, (World Scientific, Singapore), 1990.


\bibitem{Gradshteyn} I. S. Gradshteyn, I. M. Ryzhik, \emph{Table of Integrals, Series, and Products}, A. Jeffrey and D. Zwillinger (eds), (Academic Press,
San Diego), 2007.

\end{thebibliography}
\end{document}